\documentclass[aps,preprint,nofootinbib,floatfix]{revtex4-2}
\usepackage[utf8]{inputenc}
\usepackage{subfigure}
\usepackage[graphicx]{realboxes}
\usepackage{graphicx,tabularx}
\usepackage{float}
\usepackage{setspace}
\usepackage{hyperref}
\usepackage{bm}
\usepackage{multirow}
\usepackage{amsthm,amsmath,amssymb}
\usepackage{mathrsfs}
\usepackage{amsmath}
\usepackage{indentfirst}
\usepackage{tikz}
\usepackage{url}
\usepackage{footmisc}
\usepackage{amsfonts}
\usepackage{cancel}
\usepackage{color}
\usepackage{multirow}
\usepackage{slashed}
\usepackage{comment}
\usepackage{appendix}
\usepackage{soul}
\usepackage{setspace}
\usepackage{titletoc}
\usepackage{booktabs}

\begin{document}
\title{Research of Extra Charged Gauge Boson $W^{\prime}$ in Alternative Left-Right Model at Future Muon Collider}
\author{Liuxin Zhao}
\author{Honglei Li$^{a}$}
\email{sps\_lihl@ujn.edu.cn}
\author{Zhi-Long Han$^{a}$}
\email{sps\_hanzl@ujn.edu.cn}
\author{Fei Huang$^{a}$}
\email{sps\_huangf@ujn.edu.cn}
\author{Xinyi Yan} 
\affiliation{$^a$School of Physics and Technology, University of Jinan, 250022, Jinan, China}
\begin{abstract}
The study of extra charged gauge boson beyond the Standard Model has always been of great interest. Future muon colliders will have a significant advantage in discovering exotic particles. In this paper, by studying the $\mu^+  \mu^- \to W^{\prime +} W^{\prime -} \to e^+ e^- n_e \bar{n}_e$ process, we explore the properties of $W^\prime$ in the alternative left-right model.
The cross section and angular distribution of the final electron are investigated in the scenario of different $W^\prime$ mass and right-handed coupling constant. 
The forward-backward asymmetry is also an important observable to reflect the properties of $W^\prime$.
We provide a method to effectively suppress the background processes. With specific kinematic cuts, the significance can reach  $5.17\sigma$ for 4.8 TeV $W^\prime$ at the collision energy of 10 TeV.

\end{abstract}
\maketitle
\newpage

\section{Introduction}
\par  The exploration of new physics is an active field in particle physics, aiming to go beyond the limitations of the Standard Model (SM) and solve mysteries such as dark matter, dark energy, and the mass of neutrino, etc. Among them, supersymmetry theory (SUSY)~\cite{1,2,4} grand unified theory (GUT)~\cite{3,5,6}, extra dimension theory and related extended theories ~\cite{Senjanovic:1975rk,Liu:2022byu,He:2024dwh,Frank:2019nid,Ashry:2013loa}  are  the popular theories in the process of new physics research. 
Additional gauge particles are predicted along with the new gauge group involved.
The detection of the $W^\prime $ particle is an important goal in new physics models. Recently, the CDF experiment group at Fermilab, through precise measurements of the $W$ boson mass, found deviations from the expectations of the Standard Model~\cite{CDF:2022hxs}, which may suggest the existence of the $W^\prime $  particle or other new physics. 

The $W^\prime $ particle was proposed mainly to address some of the limitations and mysteries of the Standard Model. It is a hypothetical particle beyond the predictions of the Standard Model, resulting from new theories and models, and the properties of the $W^\prime$ particle have different predictions in different models.
The origin and background of the $W^\prime $ particle are closely related to the electroweak symmetry breaking and the Higgs mechanism. In the Standard Model, $W$ and $Z$ bosons acquire masses through the Higgs mechanism. If the $W^\prime $ particle exists, it may involve a relatively more complex mode of symmetry breaking. Therefore, the existence of the $W^\prime $ particle may point to new phenomena, such as the aforementioned GUT, extra dimension theory, which provide a possible framework to explain the properties of the $W^\prime $ particle and its interactions with other particles. The alternative left-right model is an attractive model breaking from the GUT, which provides the explanation of neutrino masses and dark matter candidates. 
We choose the alternative left-right model for the studies of $W^\prime $. In this model, the $W^\prime$ particle can allow for the existence of light masses, and its mass is influenced by the coupling constants. The ratio of $M_{W^\prime}/M_{Z^\prime}$ depends on the right-handed coupling constant $g_R$. The model introduces a $S$ symmetry, which provides theoretical stability for the newly neutral fermions and Higgs particles that can serve as dark matter candidates. At the same time, the model avoids flavor-changing neutral currents, which allows for the charged bosons to be light. One can study the interactions of $W^\prime $ with the Standard Model  particles and further explore the dark matter interactions with the $W^\prime$ particle as a mediator, and so on.
Overall, the exploration of the $W^\prime $ particle is an exciting field in particle physics. It may not only reveal clues to new physics~\cite{Lu:2023jlr,Yin:2021rlr,Gong:2014qla,Kriewald:2024cgr,Frank:2023epx} but also provide us with the key to a deeper understanding of the universe. In the coming years, we are looking forward to further exploration of the $W^\prime $ particle in the future colliders.

In this paper, we firstly introduce the relevant research content related to $W^\prime $ particles. In the second section, we focus on  the model building of this study and the current mass constraints of $W^\prime $ particle. In the third section, we investigate the process $\mu^+ \mu^-\to W^{\prime +} W^{\prime -} \to e^+ e^- n_e \Bar{n}_e $  at the future muon collider. Finally, a brief summary is given.

\section{$W^\prime $ in the Alternative Left-Right Model and the constraints}

The research of grand unified theory (GUT) is widely popular, originating from the symmetry breaking of the $E_6$ group~\cite{Achiman:1978vg,Gursey:1975ki}.
The group of $SO(10) \times U(1)$ and $SU(3) \times SU(3) \times SU(3)$ are two of the many subgroups of the $E_6$ group, where the classical left-right symmetric model arises from the $SO(10) \times U(1)$ group structure~\cite{eff,tlr}. And the alternative left-right  model (ALRM)~\cite{Frank:2019nid,Frank:2024imi} discussed in the work is obtained by the breaking chain of the $SU(3) \times SU(3) \times SU(3)$ subgroup.
\par The quantum numbers and representations selected for the fermionic field content of the ALRM are inspired by heterotic superstring models~\cite{Frank:2019nid}. With the largest subgroup $SU(3)_C \times SU(3)_L \times SU(3)_H$ of $E_6$, there has 27 representation to display the particles
\begin{equation}
27=(3,3,1)+(\Bar{3},1,\Bar{3})+(1,\Bar{3},3)\equiv q+\Bar{q}+l.
\end{equation}

 The alternative left-right model is based on the $SU(3)_C \times SU(2)_L \times SU(2)_{R^\prime} \times U(1)_{B-L} \times U(1)_S $ gauge group.  In this context, the $SU(3)_C$ group represents the symmetry group of the strong interaction, corresponding to the strong force that binds quarks together; $SU(2)_L $ is the symmetry of the weak interaction for left-handed particles; and $SU(2)_{R^\prime}$ is the symmetry of the weak interaction for right-handed particles. Initially, the gauge and global symmetry group $SU(2)_{R^\prime} \times U(1)_{B-L} \times  U(1)_S $ of ALRM is broken to produce the hypercharge $U(1)_Y $. The symmetry breaking is achieved through the $SU(2)_{R^\prime} $ doublet of scalar fields $\chi_R$, which are charged under $ U(1)_S $. To maintain left-right symmetry, an $SU(2)_L $ doublet of scalar fields $\chi_L $ is introduced. However, unlike  $\chi_R $ ,  $\chi_L $ does not couple with $ U(1)_S $. The electroweak symmetry is further broken down to electromagnetism through a bidoublet Higgs field that carries charges under both $SU(2)_L $ and $SU(2)_{R^\prime} $, but has no $B-L$ quantum numbers. This Higgs field is non-trivial under both $SU(2)_L $ and $SU(2)_{R^\prime} $ transformations but does not affect the $B-L$ symmetry~\cite{Ashry:2013loa}. We represent it with a more intuitive process as follows
\begin{equation}
    SU(2)_{L} \times SU(2)_{R^\prime} \times U(1)_{B-L}\times U(1)_S \xrightarrow{\langle\chi_{R}\rangle} SU(2)_{L} \times U(1)_{Y} \xrightarrow{\langle\Phi\rangle,\langle\chi_{L}\rangle} U(1)_{EM}.
\end{equation}
The particles of ALRM with the representation of field are listed in table~\ref{tb1}. Under the $S$ symmetry, there are some odd $S$ symmetry particles: the scotinos $n_L$,$n_R$, the exotic quarks $d_L^{\prime}$, $d_R^{\prime}$, the gauge boson $W^\prime$, and the scalar particles. The Yukawa Lagrangian of ALRM is written as
 \begin{equation}
     \mathcal{L}_{Y} = \bar{Q}_{L} Y^{q} \tilde{\Phi} Q_{R} + \bar{Q}_{L} Y_{L}^{q} \chi_{L} d_{R} + \bar{Q}_{R} Y_{R}^{q} \chi_{R} d_{L}^{\prime} + \bar{L}_{L} Y^{\ell} \Phi L_{R} + \bar{L}_{L} Y_{L}^{\ell} \tilde{\chi}_{L} \nu_{R} + \bar{L}_{R} Y_{R}^{\ell} \tilde{\chi}_{R} n_{L} + h.c.,
 \end{equation}
 where Y is a $3\times3$ matrix of Yukawa couplings, with subscripts $L$ and $R$ representing the left-handed and right-handed coupling matrices respectively, and superscripts $q$ and $l$ representing the quarks and leptons in the model. The vacuum expectation values (vevs) of the scalar fields is expressed as
 \begin{equation}
     \langle\Phi\rangle = \frac{1}{\sqrt{2}}
\begin{pmatrix}
0 & 0 \\
0 & k
\end{pmatrix},
\quad
\left\langle\chi_{L}\right\rangle = \frac{1}{\sqrt{2}}
\begin{pmatrix}
0 \\
v_{L}
\end{pmatrix},
\quad
\left\langle\chi_{R}\right\rangle = \frac{1}{\sqrt{2}}
\begin{pmatrix}
0 \\
v_{R}
\end{pmatrix}.
 \end{equation}
 The masses of the gauge bosons are generated through the left-right symmetry breaking the Higgs mechanism. The masses of the charged bosons  are the follows 
 \begin{equation}
     M_{W} = \frac{1}{2} g_L \sqrt{k^2 + v_L^2} \equiv \frac{1}{2} g_L v, \quad \text{and} \quad M_{W^\prime} = \frac{1}{2} g_R \sqrt{k^2 + v_R^2} \equiv \frac{1}{2} g_R v^\prime . 
     \label{Mw}
 \end{equation}
 However, due to $\langle\phi_1^0\rangle = 0$, the produced charged bosons will not mix, which means that the $W^{\prime\pm}$ and $W^\pm$ particles will maintain their respective characteristics and will not convert into each other. The mass squared matrix of the gauge bosons is
 \begin{equation}
(\mathcal{M}_V^0)^2 = \frac{1}{4}
\begin{pmatrix}
g_{B-L}^2 (v_L^2 + v_R^2) & -g_{B-L} g_L v_L^2 & -g_{B-L} g_R v_R^2 \\
-g_{B-L} g_L v_L^2 & g_L^2 v^2 & -g_L g_R k^2 \\
-g_{B-L} g_R v_R^2 & -g_L g_R k^2 & g_R^2 (v')^2
\end{pmatrix},
\end{equation}
where $g_L,g_R$ and $g_{B-L}$ are the gauge coupling constants of the $SU(2)_L,SU(2)_{R^\prime}$ and $U(1)_{B-L}$.

\begin{table}[H]
\begin{center}
\setlength{\abovecaptionskip}{6pt}
\setlength{\belowcaptionskip}{0pt}
\begin{tabular}{ |c |c |c |}
\hline
\hline
 Fields & $SU(3)_C \times SU(2)_L \times SU(2)_{R^\prime} \times U(1)_{B-L}$ & S \\
\hline 
 Fermions &  &    \\
$Q_L=\begin{pmatrix} u \\ d \end{pmatrix}_L$ & $\begin{pmatrix} 3,2,1,+\frac{1}{6} \end{pmatrix}$ & $0$\\

$Q_R=\begin{pmatrix} u \\ d^\prime \end{pmatrix}_R$ & $\begin{pmatrix} 3,1,2,+\frac{1}{6} \end{pmatrix}$ & $-\frac{1}{2}$\\
 
$d^\prime_L $ & $\begin{pmatrix} 3,1,1,-\frac{1}{3} \end{pmatrix}$ & $-1$\\

$d_R $& $\begin{pmatrix} 3,1,1,-\frac{1}{3} \end{pmatrix}$ & 0\\

$L_L=\begin{pmatrix} \nu \\ e \end{pmatrix}_L$ & $\begin{pmatrix} 1,2,1,-\frac{1}{2} \end{pmatrix}$ & $0$\\

$L_R=\begin{pmatrix} n \\ e \end{pmatrix}_R$ & $\begin{pmatrix} 1,1,2,-\frac{1}{2} \end{pmatrix}$ & $+\frac{1}{2}$\\

$n_L $ & $\begin{pmatrix} 1,1,1,0 \end{pmatrix}$ & $+1$\\

$\nu_R $& $\begin{pmatrix} 1,1,1,0 \end{pmatrix}$ & 0\\
\hline
Higgs & & \\
$\Phi=\begin{pmatrix} \phi_1^0  \phi_1^+ \\ \phi_2^-  \phi_2^0 \end{pmatrix}_R$ & $\begin{pmatrix} 1,2,2^*,0 \end{pmatrix}$ & $-\frac{1}{2}$\\

$X_L=\begin{pmatrix} \chi_L^+ \\ \chi_L^0 \end{pmatrix}_L$ & $\begin{pmatrix} 1,2,1,+\frac{1}{2} \end{pmatrix}$ & $0$\\

$X_R=\begin{pmatrix} \chi_R^+ \\ \chi_R^0 \end{pmatrix}_L$ & $\begin{pmatrix} 1,1,2,+\frac{1}{2} \end{pmatrix}$ & $+\frac{1}{2}$\\
\hline

\end{tabular}
\caption{Field content of the ALRM and respective quantum numbers.}
\label{tb1}
\end{center}
\end{table}

In recent years, many research groups (such as the ATLAS Collaboration and the CMS Collaboration) have been actively exploring the experimental constraints on the mass of the $W^\prime $ particle, and have achieved a lot of research results~\cite{ATLAS1,ATLAS3,CMS2,CMS3,1807.11421,CMS:2022krd,CMS:2021dzb,ATLAS:2024tzc,ATLAS:2019isd}. Below are some of the latest research findings on the mass constraints of the $W^\prime $:

1.Bounds from CMS: The CMS Collaboration has searched for $W^\prime $ and heavy neutrino $N_R$ with the signature of observing high transverse masses of the $\tau-$lepton and missing transverse momentum at the LHC with $\sqrt{s}=13$ TeV, corresponding to the integrated luminosity of $35.9\ fb^{-1}$. The mass bounds of $W^\prime$  is $M_{W^\prime}\geq 4.0 $ TeV, with the right-handed coupling equal to SM coupling~\cite{1807.11421}. But depending on the coupling in the non-universal $G(221)$ model, the bounds of heavy $W^\prime$ bosons is $M_{W^\prime}\geq 1.7\sim3.9 $ TeV~\cite{1807.11421}.
Within the framework of the Sequential Standard
Model, a $W^\prime $ boson with mass less than 5.3 TeV is excluded from the combined results of electron decay channels at the LHC with $\sqrt{s}=13$ TeV~\cite{CMS:2022krd}.
The CMS collaboration utilizing data from Run 2 at the LHC has obtained constraints on the mass of the $W^\prime$ boson with Left-Right Symmetric Model at $\sqrt{s}=13 $ TeV. 
 For the assumption of the right-handed neutrino masses equal to half the $W^\prime$ mass, the mass of $W^\prime$ is excluded at 95\% confidence level up to 4.7 TeV and 5.0 TeV for the electron and muon channels~\cite{CMS:2021dzb}.

2.Bounds from ATLAS: 
The ATLAS collaboration has searched for $W^\prime $ with observing at high transverse masses of the $\tau-$lepton and missing transverse momentum at the LHC with $\sqrt{s}=13 $ TeV, corresponding to the integrated luminosity of $138\ fb^{-1}$. The present mass bounds of $W^\prime$ is $M_{W^\prime}\geq 5.0 $ TeV in Sequential Standard Model, with the right-handed coupling equal to SM coupling~\cite{ATLAS:2024tzc}.
For the assumption of the coupling in the non-universal coupling $g_L\neq g_R$, the mass of $W^\prime$ is excluded at 95\% confidence level up to $3.5 \sim 5.0$ TeV for the $\tau$ channel in the Sequential Standard Model.
Corresponding to the integrated luminosity of $80\ fb^{-1}$, the mass of $W^\prime$ is excluded at 95\% confidence level up to 4.8 TeV and 5.0 TeV for the electron and muon channels in the Left-Right Symmetric Model~\cite{ATLAS:2019isd}. 

  \begin{table}[H]
\begin{center}
\setlength{\abovecaptionskip}{6pt}
\setlength{\belowcaptionskip}{0pt}
\begin{tabular}{ |c|c|c|c|}
\hline
\multirow{3}{*}{CMS} & Sequential\  Standard\  Model~\cite{CMS:2022krd} & $W^\prime\to e^-\nu $ & $M_{W^\prime}\geq 5.3$TeV \\
\cline{2-4}

\multirow{3}{*}{} &\multirow{2}{*}{ Left-Right\ Symmetric \   Model~\cite{CMS:2021dzb}}  & $W^\prime\to e^- N$ & $M_{W^\prime}\geq 4.7$TeV\\
\cline{3-4}

\multirow{3}{*}{} &\multirow{2}{*}{}  & $W^\prime\to \mu^- N$ & $M_{W^\prime}\geq 5.0$TeV\\
\cline{3-4}
\hline
\multirow{4}{*}{ATLAS} & Sequential\  Standard \ Model($g_L=g_R$)~\cite{ATLAS:2024tzc} & $W^\prime\to \tau^-\nu $ & $M_{W^\prime}\geq 5.0$TeV \\
\cline{2-4}

\multirow{4}{*}{} & Sequential\  Standard \ Model($g_L\neq g_R$)~\cite{ATLAS:2024tzc} & $W^\prime\to \tau^-\nu$ & $M_{W^\prime}\geq 3.5\sim5.0$TeV\\
\cline{2-4}

\multirow{4}{*}{} & \multirow{2}{*}{Left-Right\ Symmetric \  Model~\cite{ATLAS:2019isd}}  & $W^\prime\to e^- n_e$ & $M_{W^\prime}\geq 4.8$TeV\\
\cline{3-4}
\multirow{4}{*}{} &  \multirow{2}{*}{} & $W^\prime\to \mu^- n_{\mu}$ & $M_{W^\prime}\geq 5.0$TeV\\
\cline{3-4}
\hline

\end{tabular}
\caption{The mass constraints for $W^\prime$ from LHC.}
\label{Exper}
\end{center}
\end{table}

The constraints on the $W^\prime$ particle mentioned above come from the analysis of LHC data by various experimental groups using different models. The value of right-handed coupling constant $g_R$ has a significant impact on the constraints of $W^\prime $ particle mass. When the value of $g_R$ is relatively small, the lower limit of $W^\prime $ mass might be smaller. In the ALRM adopted in this paper, the allowed range of $g_R$ for the coupling with right-handed gauge bosons is $0.37 \sim 0.765$, where the lower limit of the allowed range comes from the theoretical restriction that $g_R/g_L$ must be larger than $tan\theta_W$~\cite{Brehmer:2015cia,Dev:2016dja}, and the upper limit comes from the phenomenological specification of $g_R$~\cite{Frank:2019nid}.

\section{\texorpdfstring{$W^{\prime}$}{} boson production and decay at future muon collider }
\par In this chapter, we will investigate the properties of the $W^\prime$ with the pair production through different processes at the future $\mu$ collider. The $\mu$ collider is a proposed lepton collider with great potential for the future, and the related contents can be found in references~\cite{mu1,mu2,mu3,FCC:2018evy}.
 We have chosen the process $\mu^+ \mu^- \to W^{\prime +} W^{\prime -}  \to e^+ e^- n_e \Bar{n}_e $ as our main subject in this study. Before delving into the specific process, we firstly explore all the decay branching ratios of the $W^\prime $ particle within this model, which will help us identify the main decay channels of the $W^\prime$ and analyze them to explore the various properties of it. 
\subsection{The decay of \texorpdfstring{$W^{\prime}$}{} boson}
\par
\begin{figure}[tbh] 
\centering
\includegraphics[width=8cm,height=6cm]
{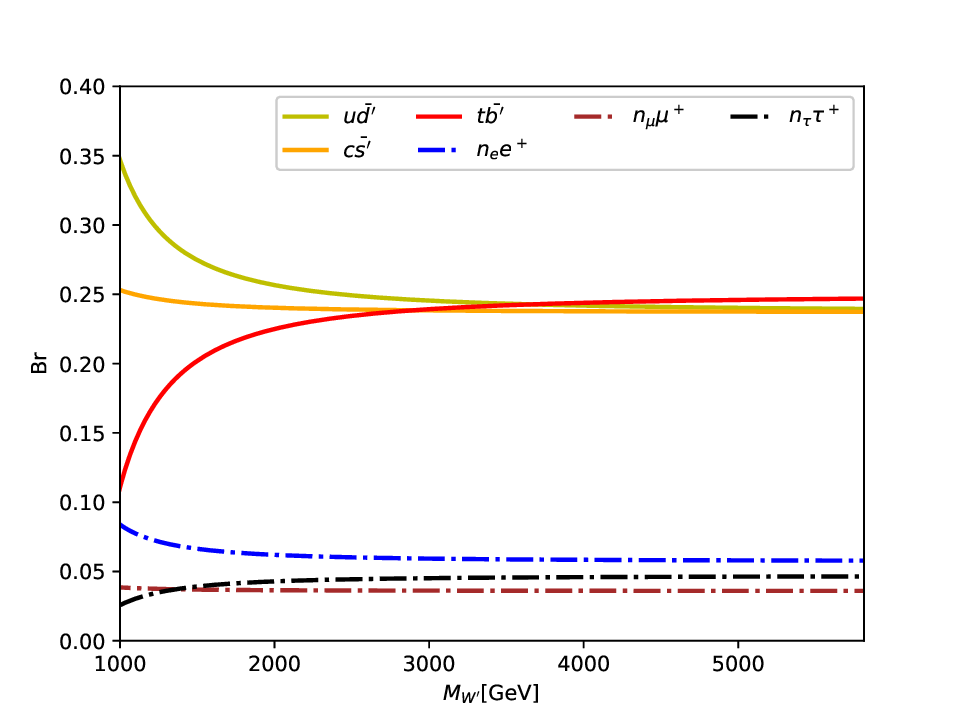}
\caption{Decay branching ratios of  $W^\prime$.}
\label{BR_part}
\end{figure}

 
In ALRM, similar to the Standard Model, gauge bosons can decay into leptons or quarks. Due to the presence of a larger number of new particles in this model, the decay channels of the $W^\prime $ boson are more complicated than the gauge particles in the Standard Model.  The main decay channels of $W^\prime $ particle are included in figure~\ref{BR_part}. 
Due to the $W^\prime $  particle's large mass, it can decay into a wider array of final states. The masses of the new particles involved are set as $M_{d^\prime ,n_e}=300 $ GeV, $M_{s^\prime ,n_\mu}=500 $ GeV, $M_{t^\prime ,n_\tau}=700 $ GeV.

\par  The coupling strength of $W^\prime $ particle decaying into all generations of leptons ($e$, $\mu$, $\tau$) are set the same for simplicity in this study. When the $W^\prime$ particle mass is less than 2 TeV, the decay branching ratio of the first generation light quarks is the largest. As the increasing of $W^\prime$ particle mass, the decay branching ratio of the third generation are raising.
As the mass of $W^\prime$ larger than 3 TeV, the decay branching ratios of quark channels tend to the similar values around 0.25. 
At the same times the decay branching ratios of electron channel is about 0.06, and the muon channel and tau channel are slightly smaller than 0.05.


 Actually, the decay process of $W^\prime$ also involves flavor-changing processes, but their decay branching ratios are less than 0.01 in our studied parameter spaces. Thus we are not plotted them in figure~\ref{BR_part}. The interesting studies on the flavor-changing processes can be found in ~\cite{flavor,flavor1,Han:flavor}, which is out of our reach in this work.

\subsection{ The process of \texorpdfstring{$\mu^+ \mu^-\to W^{\prime +}W^{\prime -} \to e^-e^+ n_e\bar{n}_e $}{}  }
\par We focus on the process of $W^\prime $ decay into a pair of light leptons, i.e., $\mu^+\mu^-\to W^{\prime +} W^{\prime -} \to e^+ e^- n_e \Bar{n}_e $.
Although the branching ratios for quark final states are significantly larger than those for lepton final states in the previous studies, we have chosen the latter as the target process in this part for the following reasons. Firstly, the final charged leptons as free particles can more easily measured than the quarks at colliders. Secondly, the right-handed neutrinos produced can be considered as candidates for dark matter, and it leads the large missing transverse  momentum as a signature. Last but not least, in our chosen decay chains it includes the couplings of $W^\prime$ and leptons without the quarks, which is benefit for the study of $W^\prime$ and leptons interactions. 

\subsubsection{Cross Section}
\begin{figure}[tbh] 

\subfigure[]{\includegraphics[width=5cm,height=3cm]
{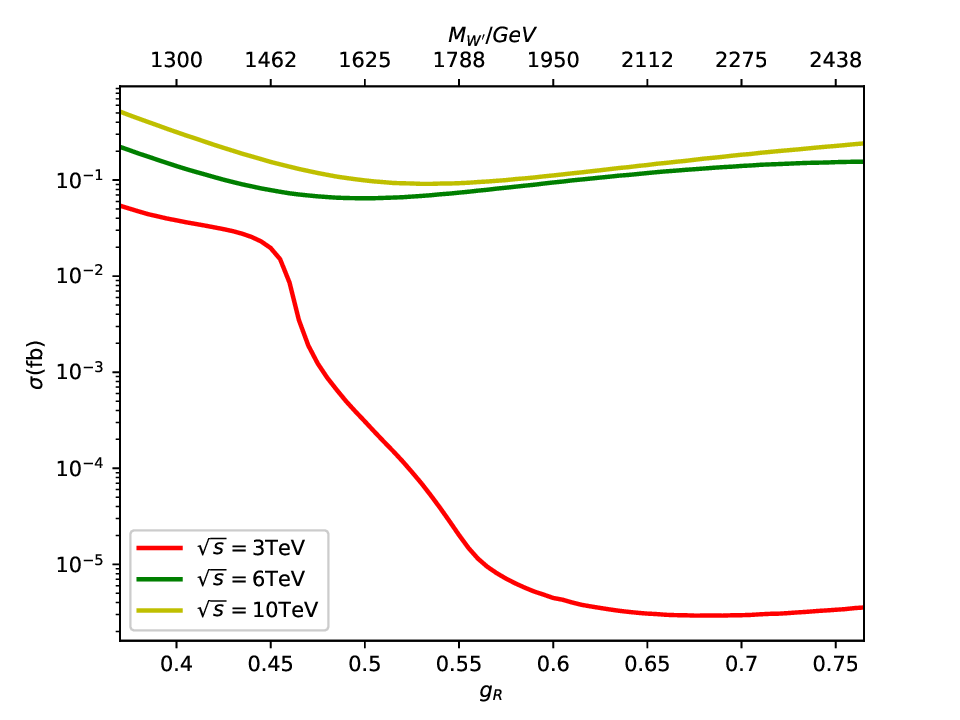}}
\subfigure[]{\includegraphics[width=5cm,height=3cm]
{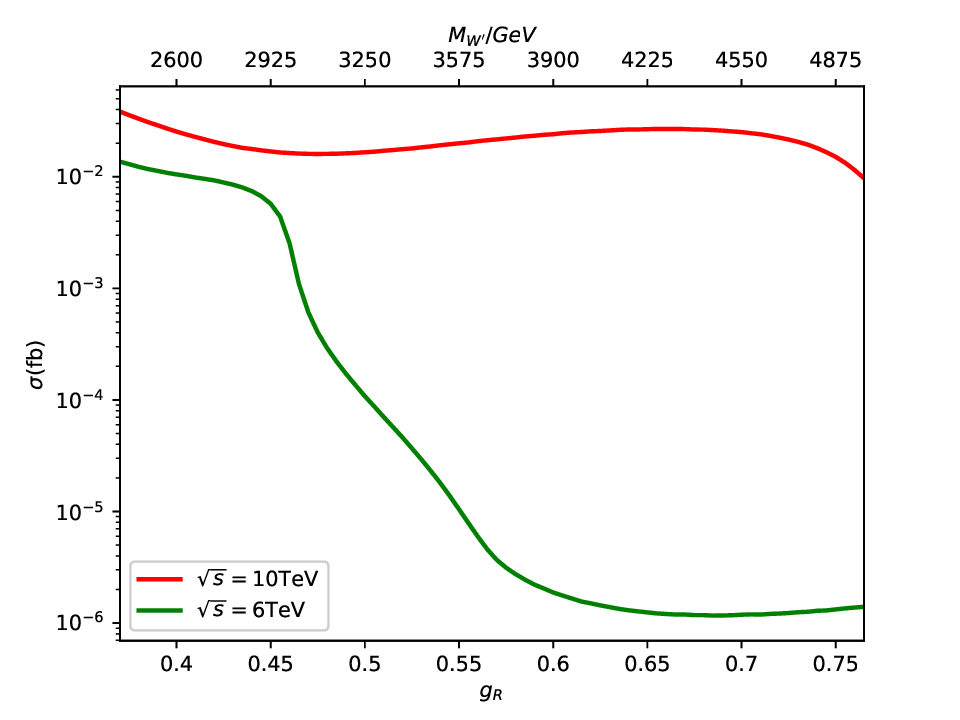}}
\subfigure[]{\includegraphics[width=5cm,height=3cm]
{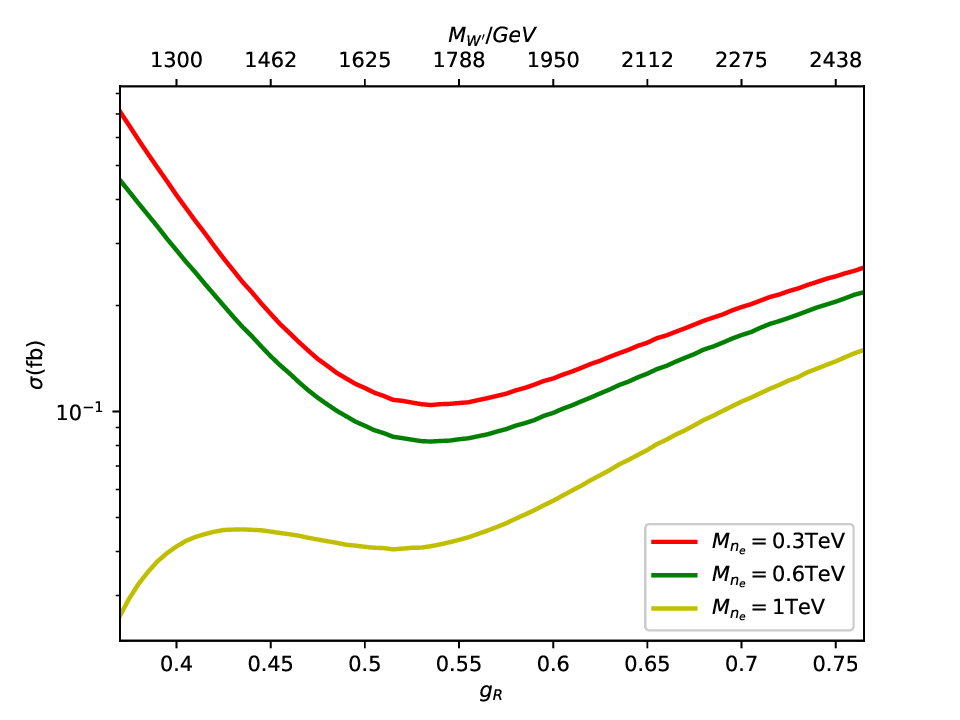}}

\caption{cross section for the process $\mu^+  \mu^- \to W^{\prime +} W^{\prime -} \to e^+ e^- n_e \bar{n}_e$ versus different couplings in figure (a) and (b). And figure (c) shows the cross section of the same process versus different $M_{n_e}$ with $\sqrt{s}=10$TeV.}
\label{CS_Lepton}
\end{figure}

\par Figure~\ref{CS_Lepton}  shows the cross section distribution of $\mu^+  \mu^- \to W^{\prime +} W^{\prime -} \to e^+ e^- n_e \bar{n}_e$ process with the right-handed coupling constant $g_R$. The plot in figure~\ref{CS_Lepton} (a) shows the cross section with different collision energies of 3, 6, and 10 TeV (represented by red, green, and yellow solid lines, respectively). In the ALRM model, the mass of the $W^\prime $ boson is determined  by the coupling constant $g_R$ and the vacuum expectation value according to equation~\eqref{Mw}, so when the vacuum expectation value is fixed, there is a one-to-one correspondence between the right-handed coupling constant $g_R$ and the mass of the $W^\prime $ particle. Therefore, we have added coordinate axes for both $g_R$ and $M_{W^\prime}$ parameters in the figure. The vacuum expectation value is fixed at 6.5 TeV in this figure. When the collision energy is 3 TeV, the cross section rapidly decreases with $g_R=0.45$, where the mass of the $W^\prime $ boson is close to 1.5 TeV. Since it is approaching the threshold for $W^\prime $ pair production.  
When $g_R$ reaches its maximum of 0.765, the corresponding mass of the $W^\prime $ boson produced is about 2.4 TeV, which means the $W^\prime $ bosons can be on shell produced with collision energy of 6 or 10 TeV in the whole studied parameter spaces.  The cross section at collision energy of 6 TeV or 10 TeV exhibits a trend of first rising and then falling with the increasing of $g_R$. This phenomena are due to the dual impact of $g_R$ and the mass of $W^\prime$.

The plot in figure~\ref{CS_Lepton} (b) also shows the cross section for $\mu^+  \mu^- \to W^{\prime +} W^{\prime -} \to e^+ e^- n_e \bar{n}_e$ process at different collision energies of 6 and 10 TeV with different $g_R$. The difference between (b) and (a) lies in the fact that we have chosen different vacuum expectation values of the Higgs field, which leads to different $W^\prime$ particle masses with same coupling constant $g_R$. The vacuum expectation value is fixed at 13 TeV in  figure~\ref{CS_Lepton} (b). 
Figure~\ref{CS_Lepton} (c) shows the cross section of $\mu^+  \mu^- \to W^{\prime +} W^{\prime -} \to e^+ e^- n_e \bar{n}_e$ process at a collision energy of 10 TeV with different neutrino mass scenarios (the red, green, and yellow solid lines represent the cross sections when the neutrino masses are 0.3 , 0.6 and 1 TeV, respectively). In this plot, the vacuum expectation value also remains 6.5 TeV. When the neutrino mass is 0.3 (1) TeV, the cross section is about 0.41 (0.041)$fb$ with $g_R=0.4$ and 0.24 (0.14) $fb$ with $g_R=0.75$.
 The neutrino mass has a significant impact on the cross section in the process of leptonic decaying channels.

\begin{figure}[tbh] 
\centering
\subfigure[]{\includegraphics[width=5cm,height=3cm]
{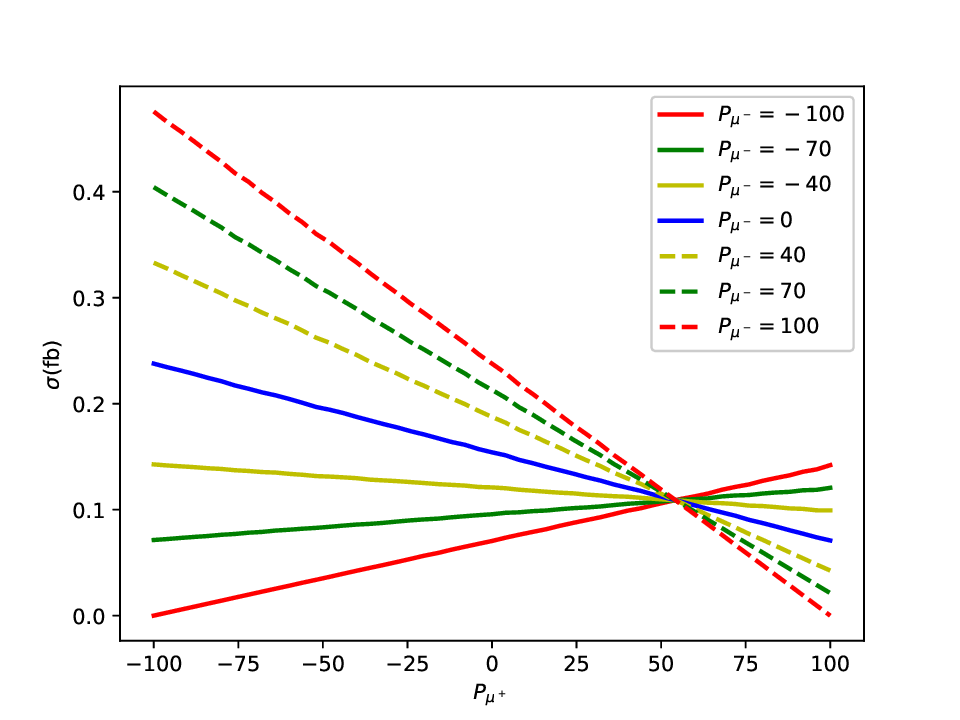}}
\subfigure[]{\includegraphics[width=5cm,height=3cm]
{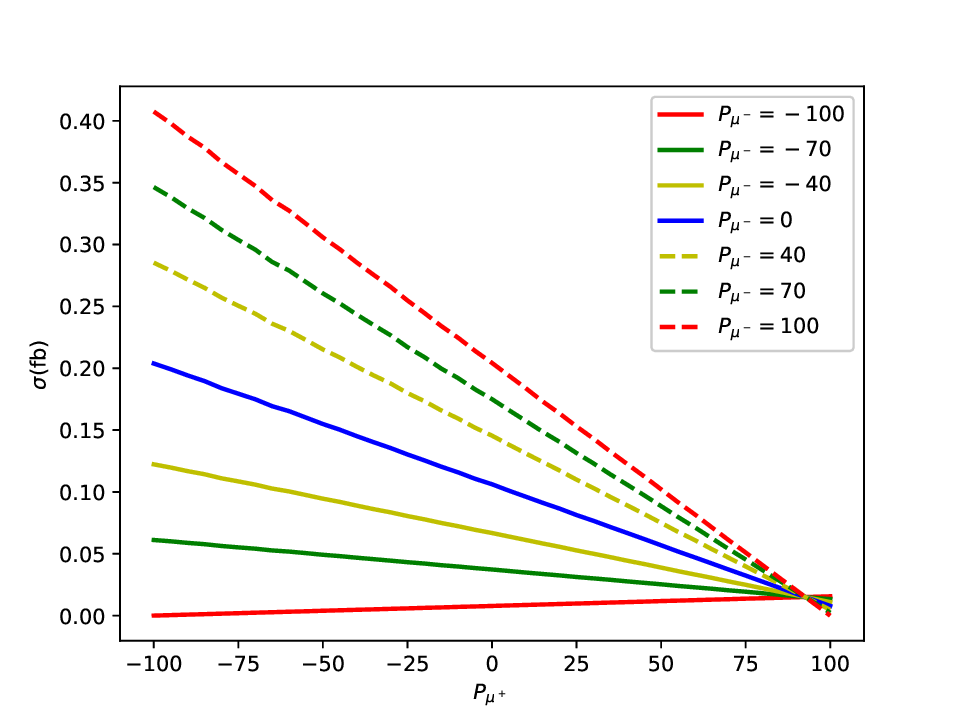}}
\subfigure[]{\includegraphics[width=5cm,height=3cm]
{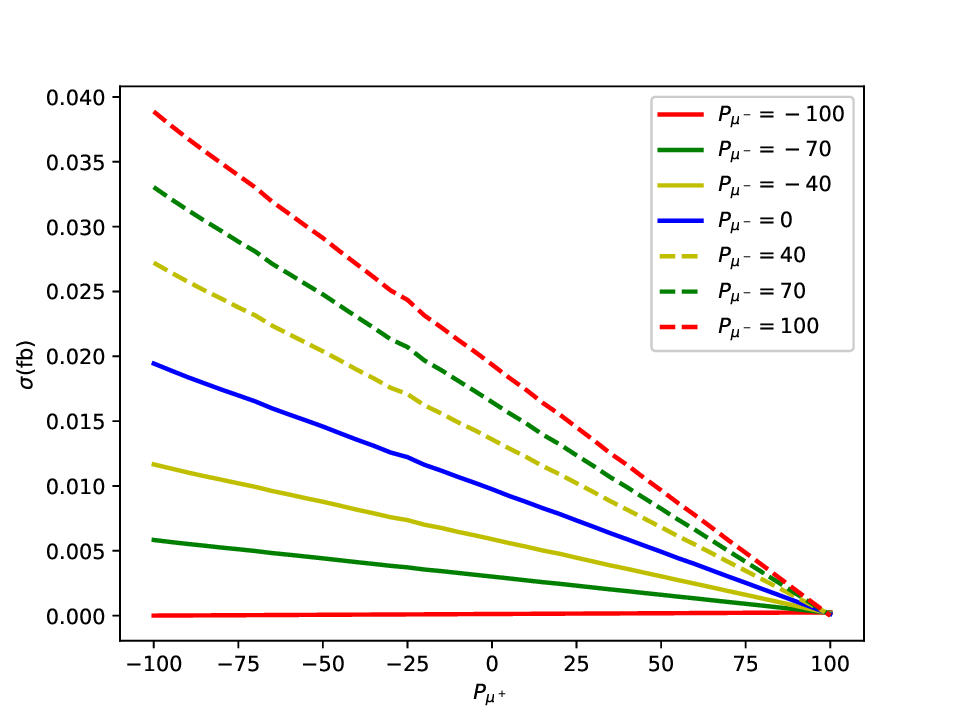}}
\caption{ Cross section distributions of $\mu^+  \mu^- \to W^{\prime +} W^{\prime -} \to e^+ e^- n_e \bar{n}_e$ process with various initial state polarization for $M_{W^\prime}=2$ TeV (a), $M_{W^\prime}=3.5$ TeV (b), and $M_{W^\prime}=5$ TeV (c) at 10 TeV muon collider.}
\label{CS_Lepton1}
\end{figure}

The polarization of the initial state particles can have a significant impact on the production of $W^\prime$ pairs. Figure~\ref{CS_Lepton1} (a)  illustrates the cross section varies with different polarization  of the initial muons through $\mu^+  \mu^- \to W^{\prime +} W^{\prime -} \to e^+ e^- n_e \bar{n}_e$ process under the condition of collision energy of 10 TeV and $m_{W^\prime}=2$ TeV. The polarization varies from fully left-handed ($P_{\mu}=-100$) to fully right-handed ($P_{\mu}=100$) with some typical values in the plots.
Each curve in the figure~\ref{CS_Lepton1} represents a fixed polarization state of the $\mu^-$ particle, with the polarization degree of the $\mu^+$ particle varying along the x-axis. 
The lines in figure~\ref{CS_Lepton1} (a) reveals an important physics phenomenon. When the polarization  of $\mu^-$ is fixed, the cross section varies with the polarization  of $\mu^+$ in different ways. 
When the polarization  of $\mu^-$ is 100, the cross section continuously decreases with the positive increase of $\mu^+$ polarization. When the polarization of $\mu^-$ is $-100$, the trend of the cross section continuously increases. Therefore, it provides the maximum and minimum values of the cross section in this process for the polarization conditions $(P_{\mu^+},P_{\mu^-})=(-100,100)$ and $(P_{\mu^+},P_{\mu^-})=(-100,-100)$. In particular, when the $\mu^-$ particles are fully polarized ($P_{\mu^-} = \pm100$), the trends in the plot become more pronounced, providing us with important clues about the characteristics of particle interactions under extreme polarization conditions. This is crucial for subsequent work in enhancing signal production.

Figure~\ref{CS_Lepton1} (b) and (c) is the same distribution as (a) but $M_{W^\prime}=3.5$ and 5 TeV.
When the mass of $W^\prime$ increases, the cross section changes significantly for $P_{\mu^-}< 0$. When $P_{\mu^-}$ equals to $-40$, the cross section for $M_{W^\prime}$ at 2 TeV increases gradually with the positive polarization of $\mu^+$, but for $M_{W^\prime}$ at 3.5 TeV, the cross section shows a decreasing trend. It is important to note that when $M_{W^\prime}$ is 3.5 and 5 TeV with $P_{\mu^-}= -100$, the cross section continuously increases with the positive polarization of $\mu^+$, but the increase is relatively small.

\subsubsection{Angular Distribution}
The angular distribution of the final particle is an important observable for the study of the properties of $W^\prime$ boson. The formula for the angular distribution is as follows,
\begin{equation}\label{eq13}
    \cos\theta=\frac{\bm{p_{f}^*} \cdot \bm{p_{i}}}{|\bm{p_{f}^*}|\cdot |\bm{p_{i}} |},
\end{equation}
where $\bm{p_{f}^*} $ and $\bm{p_{i}}$ are the three momentum of the final and initial particle, respectively. \par
\begin{figure}[tbh] 
\centering
\subfigure[]{\includegraphics[width=5cm,height=3cm]
{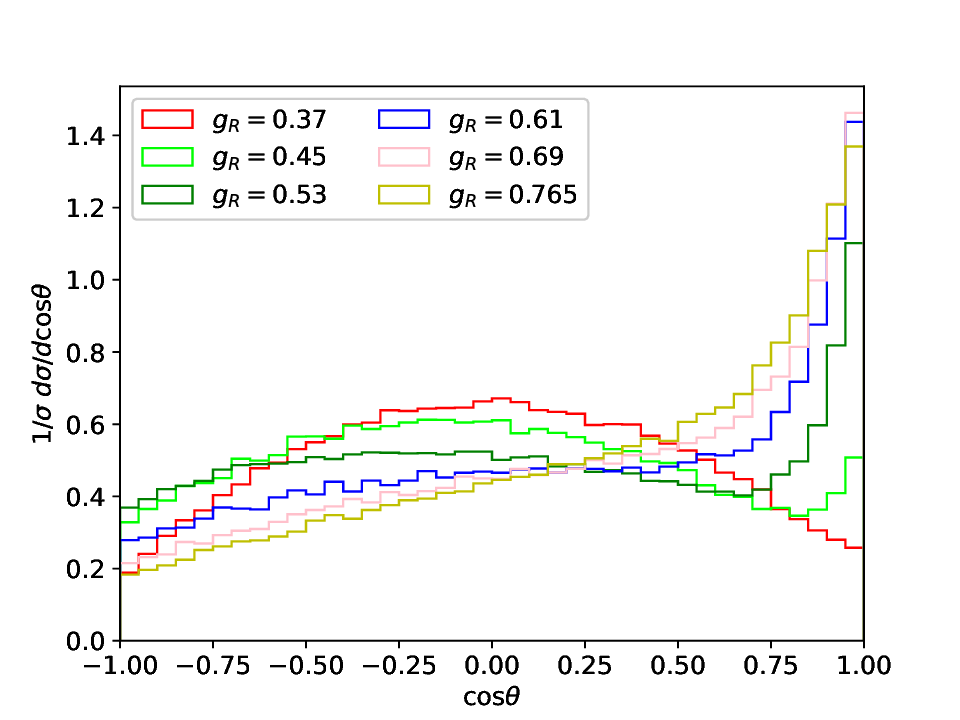}}
\subfigure[]{\includegraphics[width=5cm,height=3cm]
{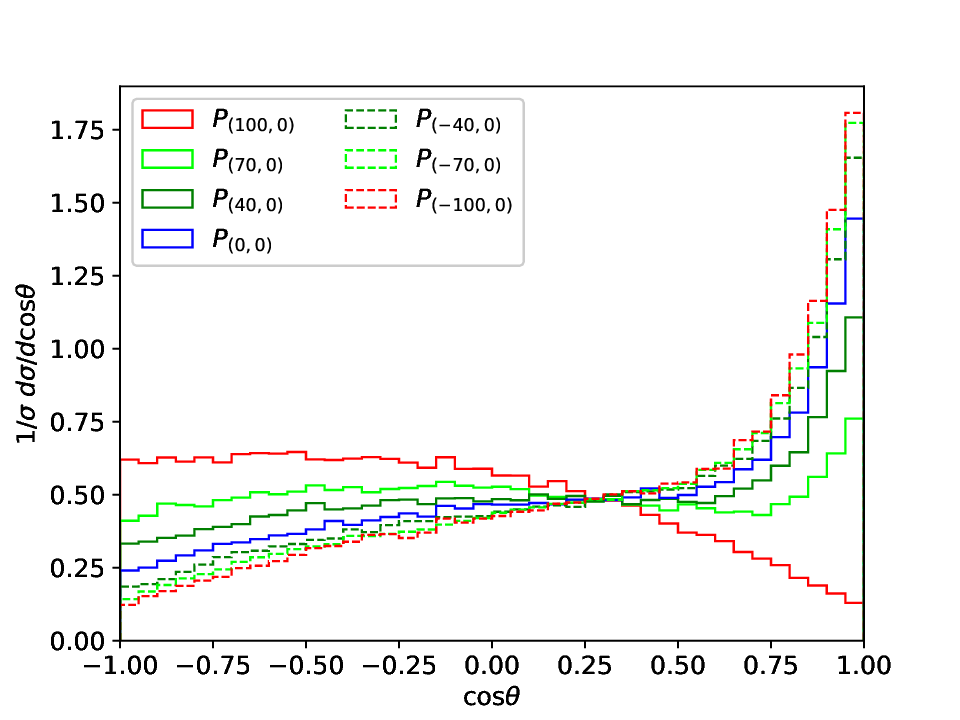}}
\subfigure[]{\includegraphics[width=5cm,height=3cm]
{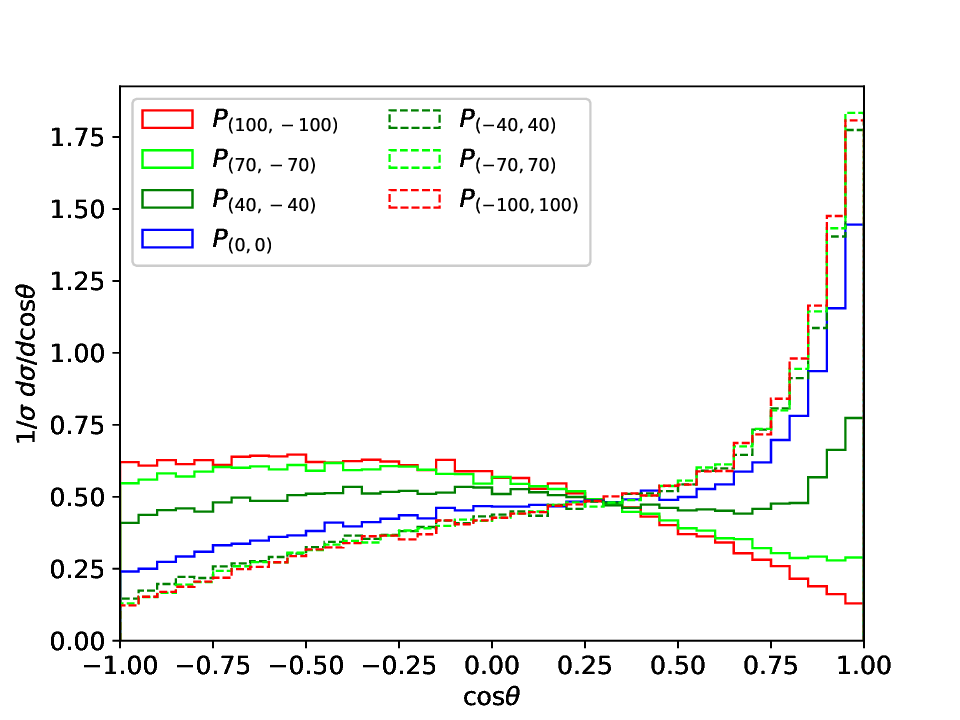}}
\subfigure[]{\includegraphics[width=5cm,height=3cm]
{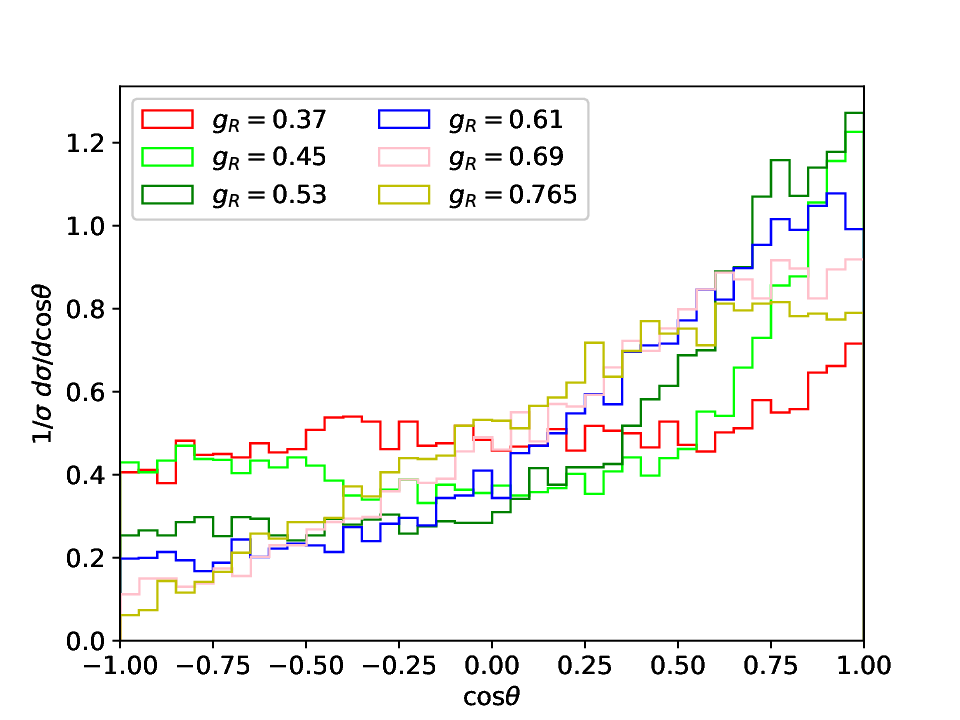}}
\subfigure[]{\includegraphics[width=5cm,height=3cm]
{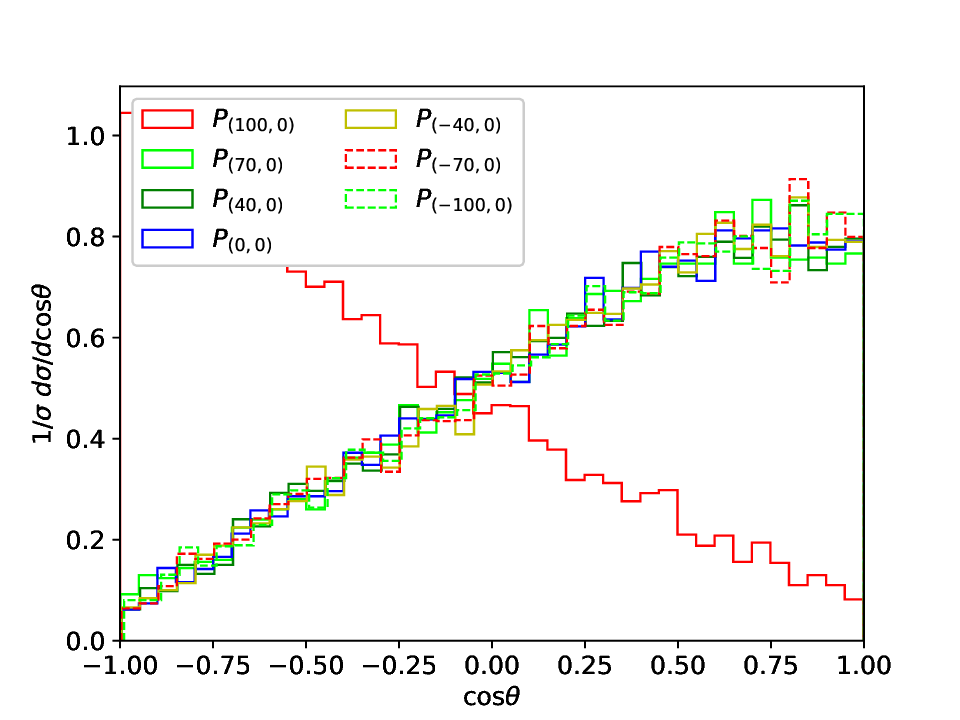}}
\subfigure[]{\includegraphics[width=5cm,height=3cm]
{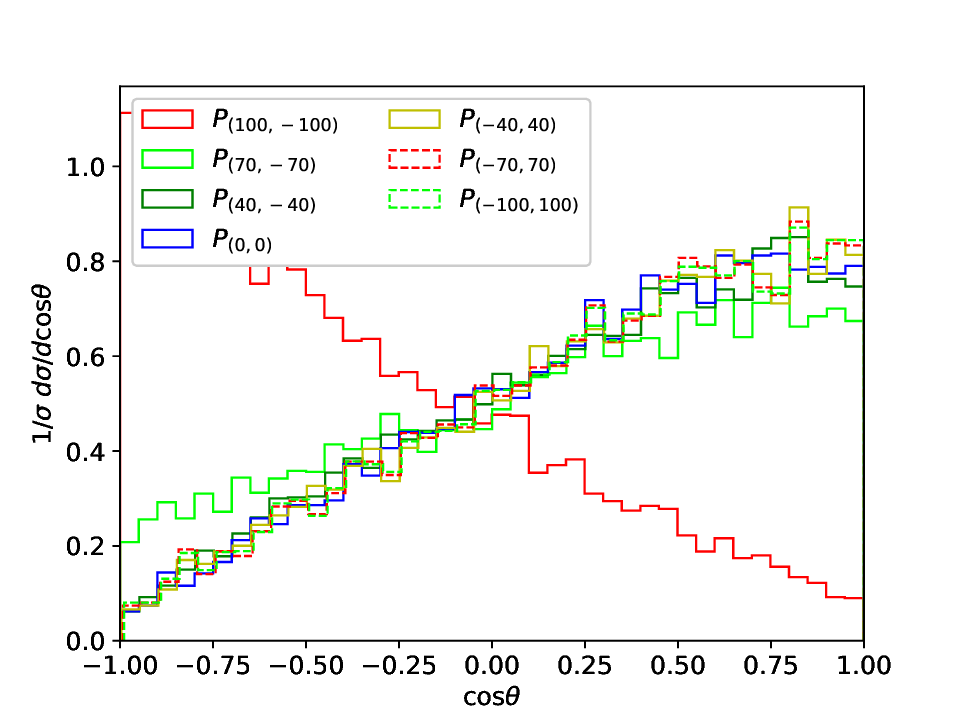}}

\caption{The angular distributions between $\mu^-$ and $e^-$ of the unpolarized and polarized  processes $\mu^+  \mu^- \to W^{\prime +} W^{\prime -} \to e^+ e^- n_e \bar{n}_e$ corresponding to the collision energy of 10 TeV. 
The vacuum expectation is 6.5 TeV in figure (a), and 13 TeV in figure (d). The fixed mass of $W^\prime$ is 2 TeV in figure (b) and (c) with $g_R=0.65$,  and $M_{W^\prime}=5$TeV, $g_R=0.765$ for (e) (f). 
}
\label{AD_Lepton}
\end{figure}

\par Figure~\ref{AD_Lepton} (a)
shows the effects of different coupling constant $g_R$ on the angular distributions between  $\mu^-$ and $e^-$ particles of process $\mu^+  \mu^- \to W^{\prime +} W^{\prime -} \to e^+ e^- n_e \bar{n}_e$, with an vacuum expectation value of 6.5 TeV. For the unpolarized initial state, the coupling constant $g_R$ has a significant impact on the angular distribution of the final state particles. When the value of the right-handed coupling constant $g_R$ is 0.37, the angular distribution is symmetric in the plot. And as the coupling constant increases, one can clearly observe an increase in the asymmetry of the angular distribution of the final state particles. When the right-handed coupling constant is 0.765, the cross section has a clear enhancement with $\cos\theta$ closing to one.

 Figure~\ref{AD_Lepton} (b) shows the impact of polarization of the initial  $\mu^-$ on the final electron angular distributions. As the polarization strength increases, the changes of the angular distributions become more pronounced. Figure~\ref{AD_Lepton} (c) shows the angular distributions resulting from polarized both initial $\mu^-$ and $\mu^+$. We can see that compared to Figure~\ref{AD_Lepton} (b), polarizing both particles to the same extent on the basis of figure~\ref{AD_Lepton} (b) does not have a significant impact on the angular distribution. 


Figure~\ref{AD_Lepton} (d) also shows the angular distributions of final state electrons with unpolarized initial particles for the vacuum expectation value of 13 TeV. Compared to Figure~\ref{AD_Lepton} (a), the increase in the vacuum expectation value makes the asymmetry of the angular distributions more noticeable, but the trend of the angular distribution as a function of $g_R$ remains unchanged. As $g_R$ increasing, the asymmetry of the angular distribution of the final state particles increases. Figure~\ref{AD_Lepton} (e) and (f) shows the angular distribution of the final state particles in the polarized condition when the $W^\prime$ mass is 5 TeV. The results show that the angular distributions tend to the same shapes with various initial polarization except for the case of $P_{\mu^+}$ extremely close to 100.

\begin{figure}[tbh] 
\centering
\subfigure[]{\includegraphics[width=5cm,height=3cm]
{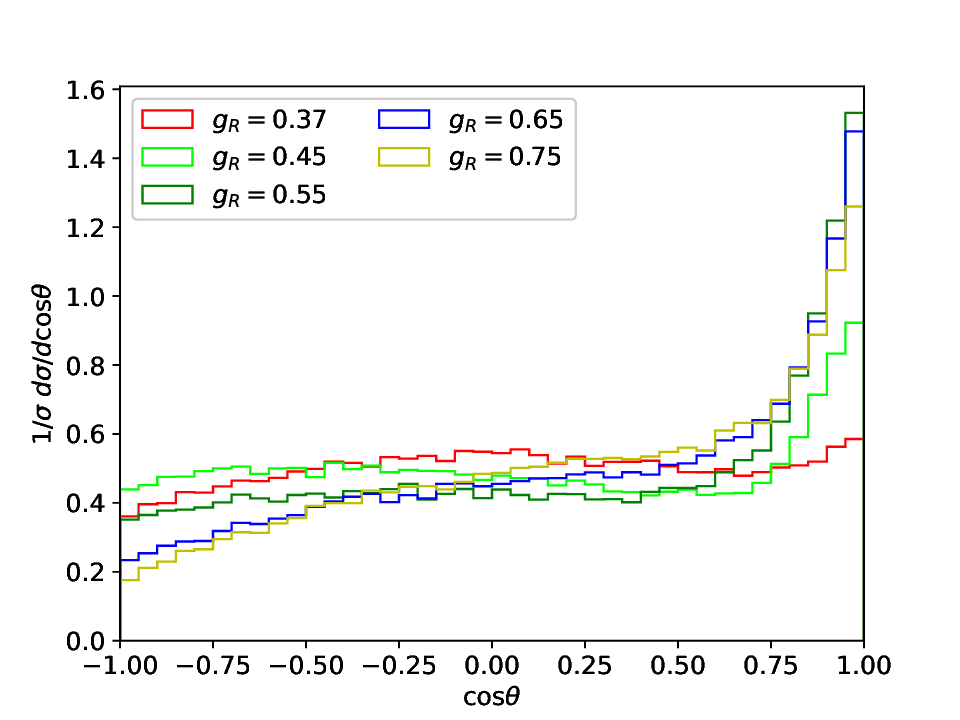}}
\subfigure[]{\includegraphics[width=5cm,height=3cm]
{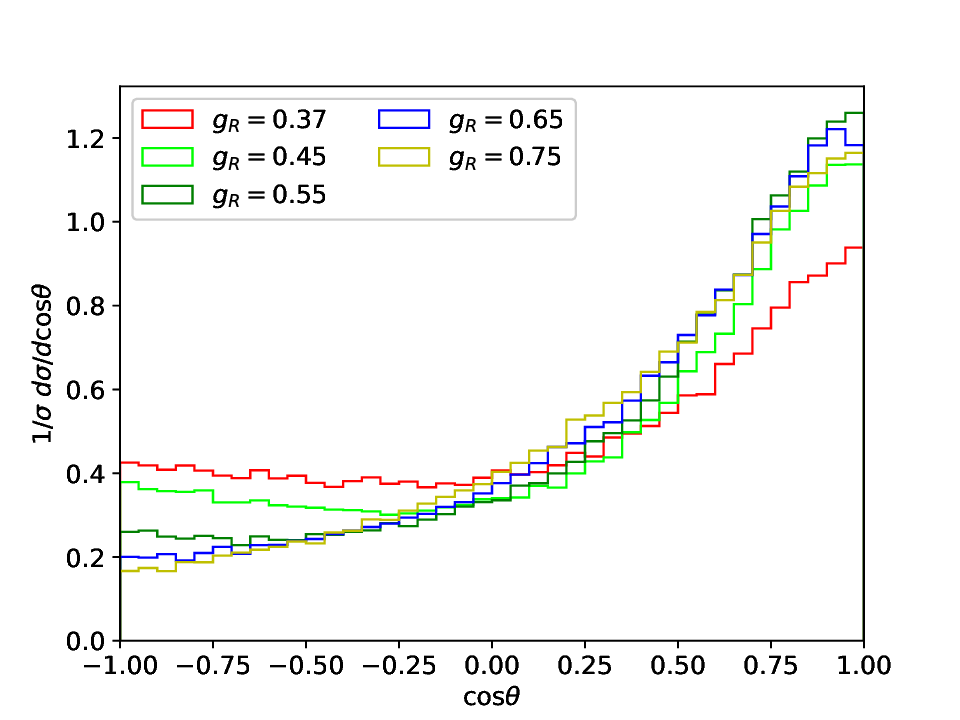}}
\subfigure[]{\includegraphics[width=5cm,height=3cm]
{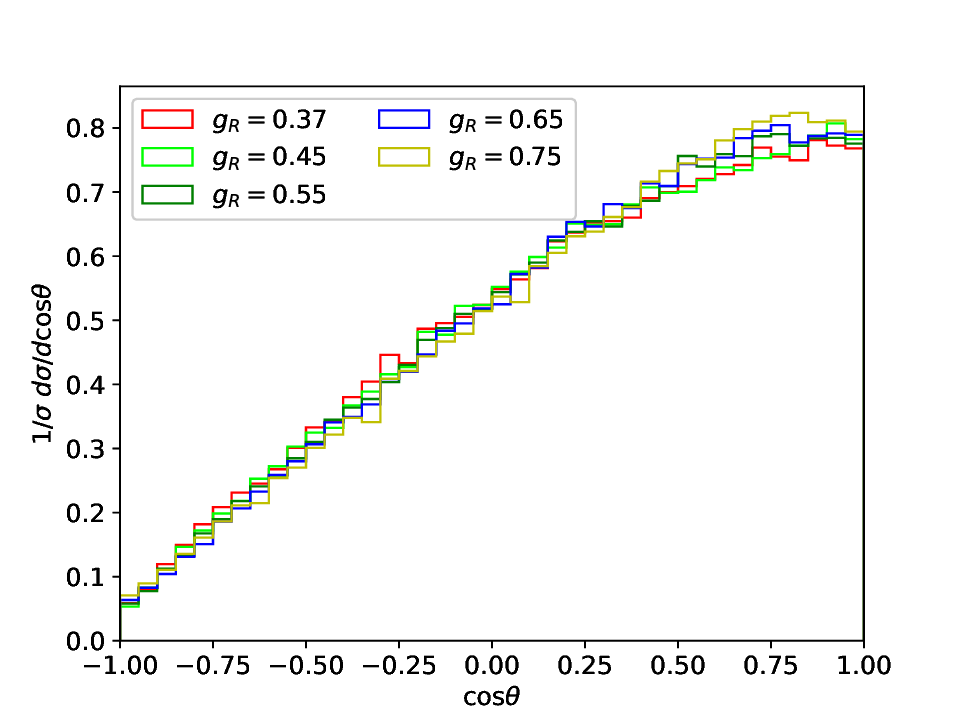}}

\caption{The angular distributions between $\mu^-$ and $e^-$ of the unpolarized processes $\mu^+  \mu^- \to W^{\prime +} W^{\prime -} \to e^+ e^- n_e \bar{n}_e$ corresponding to the collision energy of 10 TeV with  $M_{W^\prime}=$2 TeV (a),  3.5 TeV (b) and 5 TeV (c), respectively. }
\label{AD_Lepton1}
\end{figure}

The mass of $W^\prime$ has significant effects on the angular distribution of final electrons from $\mu^+  \mu^- \to W^{\prime +} W^{\prime -} \to e^+ e^- n_e \bar{n}_e$ process as shown in  Figure~\ref{AD_Lepton1}.
The mass of $W^\prime$ has been set at 2 TeV,  3.5 TeV and 5 TeV with the collision energy of 10 TeV. As the increase of $W^\prime$ mass, the asymmetry of the angular distribution becomes more obvious. Moreover, the angular distributions tend to consistency with the mass of $W^\prime$ increasing to the threshold of the $W^\prime$ pair production and the effects of different  $g_R$ values have been suppressed with the $W^\prime$ mass increasing.

Forward-backward asymmetry is a specific aspect of angular distribution that describes the difference in the emission probabilities of particles in the forward ($cos\theta \geq 0$) and backward ($cos\theta < 0$) hemispheres. We define the formula for $A_{FB}$ as follows,
\begin{equation}
    A_{FB}=\frac{\sigma(cos\theta\geq 0)-\sigma(cos\theta<0)}{\sigma(cos\theta\geq 0)+\sigma(cos\theta<0)}.
\end{equation}

\begin{table}[H]
\begin{center}
\setlength{\abovecaptionskip}{6pt}
\setlength{\belowcaptionskip}{0pt}
\begin{tabular}{ |c c|c c|c c|}
\hline 
 $g_R$ & $A_{FB}$ & $P_{(P_{\mu^+},P_{\mu^-})}$ & $A_{FB}$ & $P_{(P_{\mu^+},P_{\mu^-})}$ & $A_{FB}$  \\
\hline
$0.37$ & $0.005$ & $(100,0)$ & $-0.24$ & $(100,-100)$ & $-0.24$ \\
\hline
$0.45$ & $-0.043$ & $(70,0)$ & $0.0016$ & $(70,-70)$ & $-0.18$ \\
\hline
$0.53$ & $0.034$ & $(40,0)$ & $0.14$ & $(40,-40)$ & $0.013$ \\
\hline
$0.61$ & $0.21$ & $(-40,0)$ & $0.34$ & $(-40,40)$ & $0.39$ \\
\hline
$0.69$ & $0.31$ & $(-70,0)$ & $0.39$ & $(-70,70)$ & $0.41$ \\
\hline
$0.765$ & $0.37$ & $(-100,0)$ & $0.42$ & $(-100,100)$ & $0.42$ \\
\hline
\end{tabular}
\caption{The forward-backward asymmetry with different  conditions for $\mu^+  \mu^- \to W^{\prime +} W^{\prime -} \to e^+ e^- n_e \bar{n_e}$. The vacuum expectation value is fixed at 6.5 TeV in the first column, and $M_{W^\prime}$ is 2 TeV in the second and third column.}
\label{tb3}
\end{center}
\end{table}

\par Table~\ref{tb3} presents the forward-backward asymmetry of final electrons in the process of $\mu^+  \mu^- \to W^{\prime +} W^{\prime -} \to e^+ e^- n_e \bar{n}_e$ with various right-handed couplings and polarization of initial states. The asymmetry is 0.005 (0.37) with $g_R=0.37$ (0.765) for a fixed vacuum expectation value of 6.5 TeV. And the asymmetry varies from $-0.24$ to $0.42$ with the polarization of $P_{\mu^+}=100$ to $P_{\mu^+}=-100$. Comparing the results of the second and third columns, one can find that the inversed polarization of $\mu^-$ has slightly influence on the asymmetries. 

\subsubsection{Significance}
\par We utilize Monte Carlo simulations~\cite{MT,1995,Alwall:2014hca,Frixione:2021zdp} to simulate collisions for signal and background processes, focusing our study on the di-lepton processes. In our simulations, we pay attention to the signal process with the $W^\prime$ pair production as an intermediate state, 
\begin{equation}
    \mu^+ \mu^- \to W^{\prime +} W^{\prime -} \to e^+ e^- n_e \bar{n}_e.
\end{equation}
The main background processes include the production of $ZZ$, $WW$ particles and the VBF process directly constituted by $Z$ and neutrinos. The background processes are shown as follows
\begin{equation}
\mu^+ \mu^- \to W^+ W^- \to e^+ e^- \nu_e \bar{\nu}_e,
\end{equation}
\begin{equation}
\mu^+ \mu^- \to Z Z \to e^+ e^- \nu_e \bar{\nu}_e,\\   
\end{equation}
\begin{equation}
\mu^+ \mu^- \to Z \nu_m \bar{\nu}_m,(Z \to e^+ e^-).\\
\end{equation}
Where the left-handed or right-handed neutrinos can not be directly detected in the colliders. We treat them as the missing transverse energy. 


\begin{figure}[tbh] 
\centering
\subfigure[]{\includegraphics[width=4cm,height=3cm]
{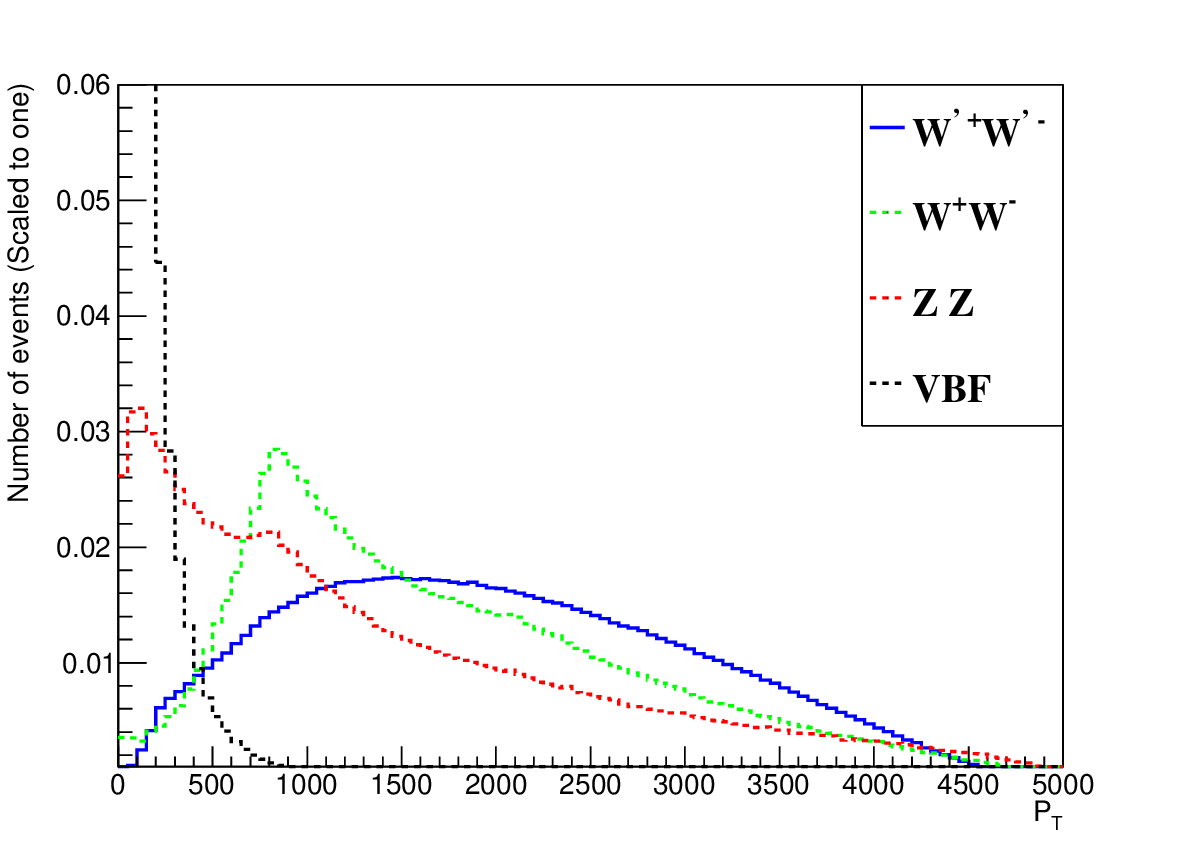}}
\subfigure[]{\includegraphics[width=4cm,height=3cm]
{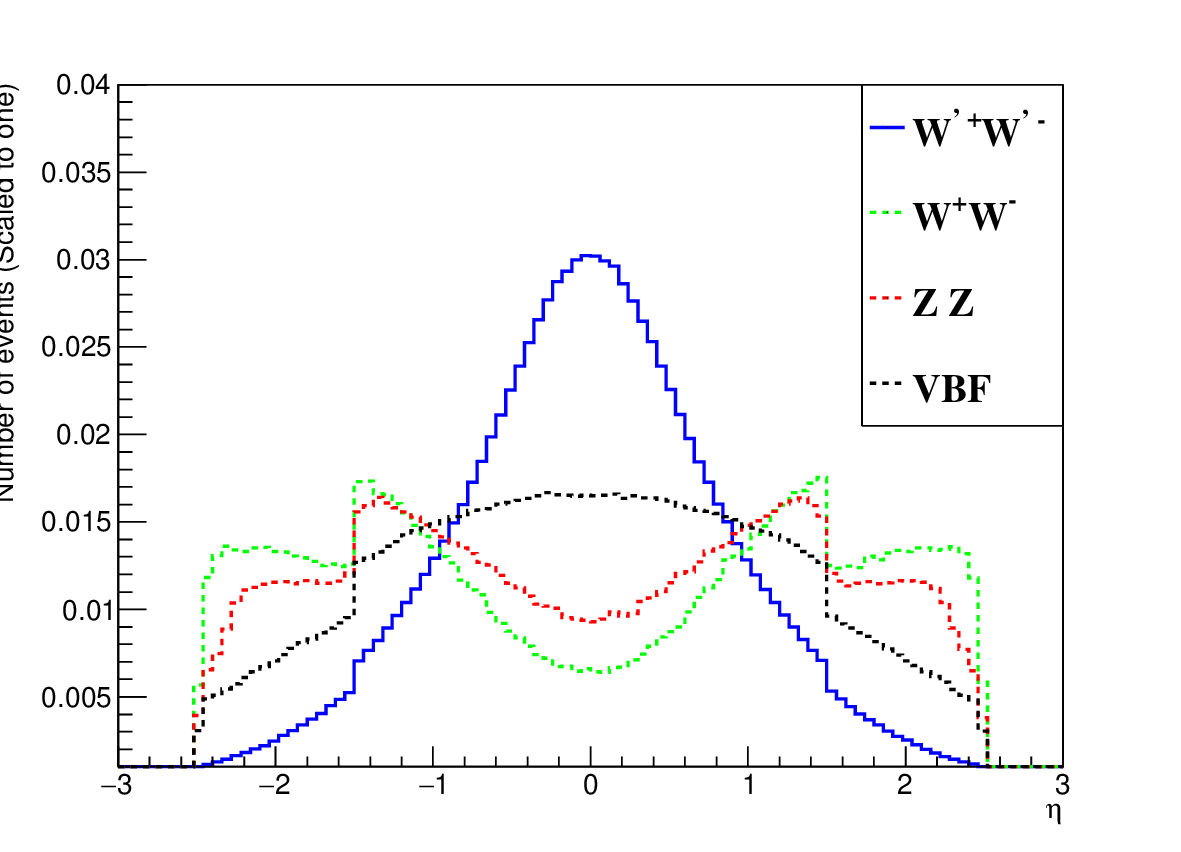}}
\subfigure[]{\includegraphics[width=4cm,height=3cm]
{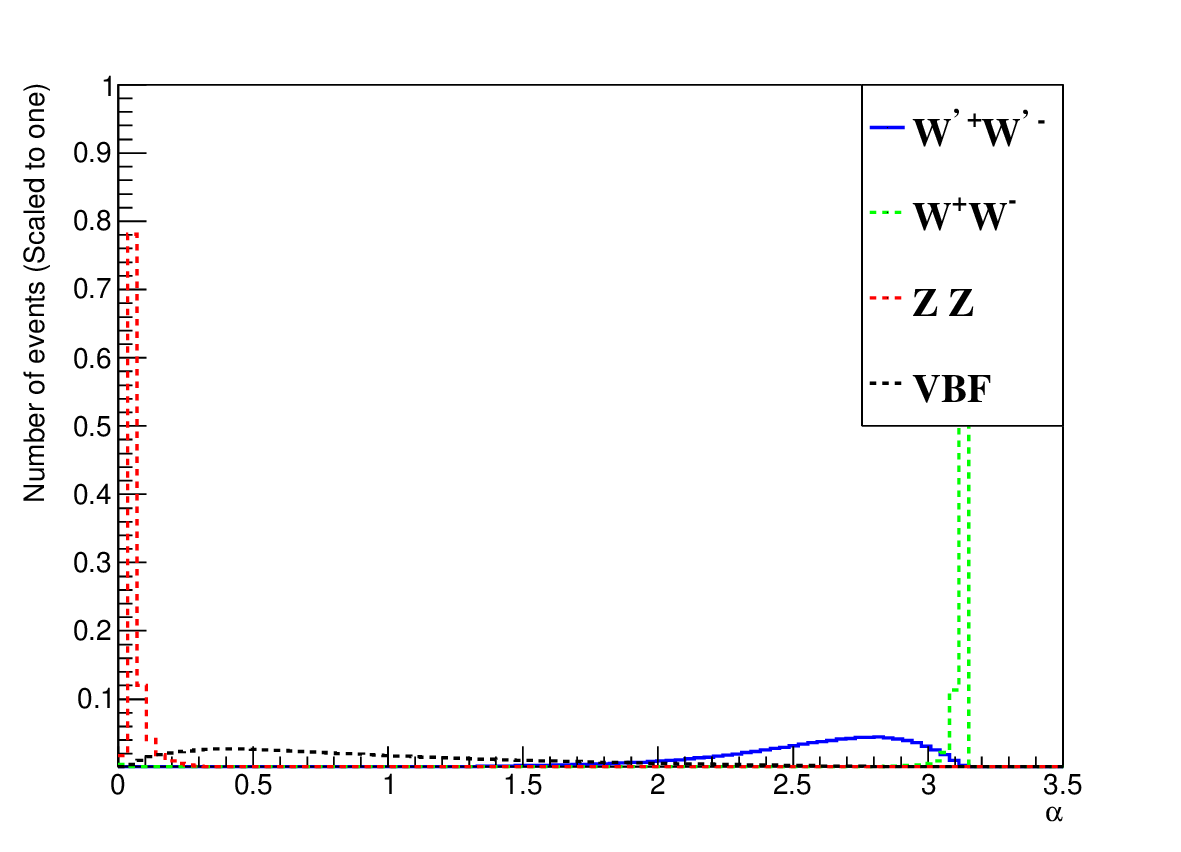}}
\subfigure[]{\includegraphics[width=4cm,height=3cm]
{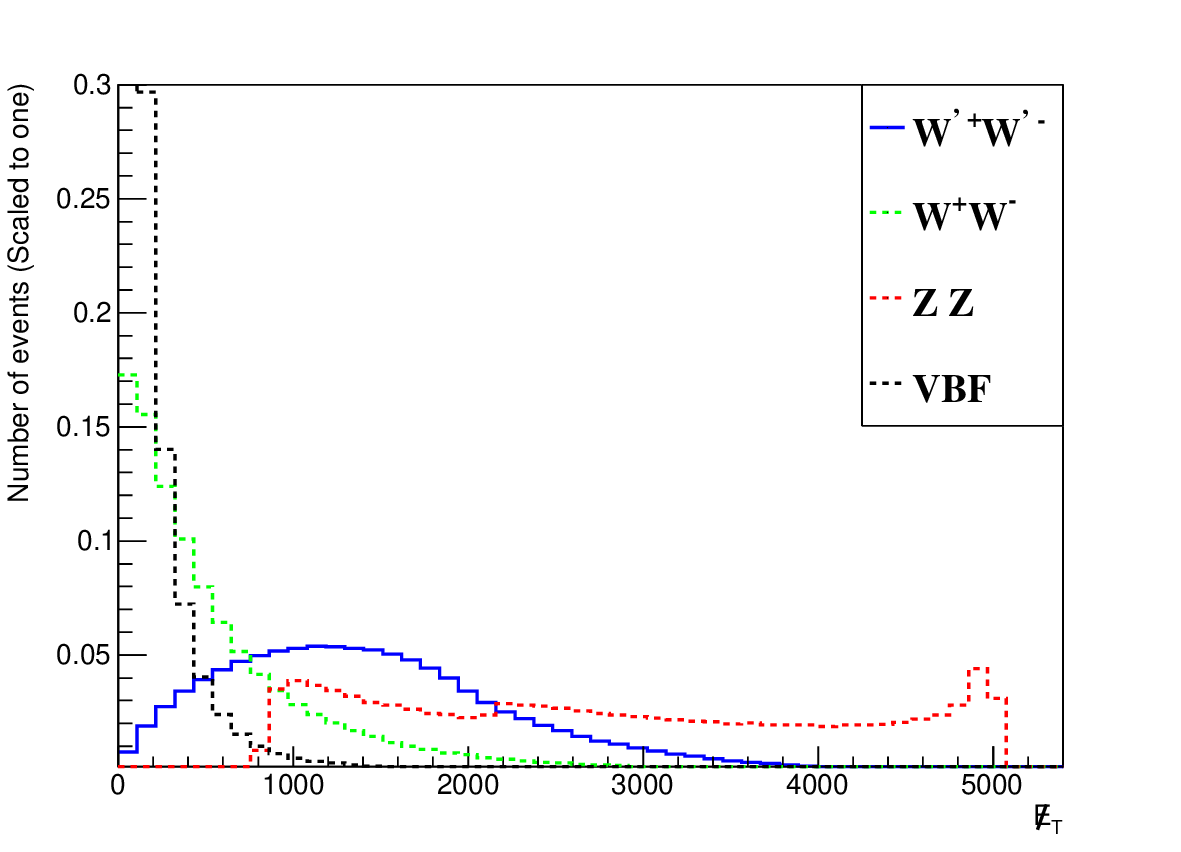}}
\caption{The kinematic distributions of the final particles for the signal and background processes before the cuts. (a), (b), (c) represent the transverse momentum $P_T$, pseudo-rapidity $\eta$, correlated angle $\alpha$ between the two charged particles in the final state and (d) is the missing transverse energy $\slashed{E}_T$.}
\label{B_cut}
\end{figure}

The signal-to-noise ratio is a measure of the strength of a signal relative to the level of background noise. Another way to compare the signal and backgrounds are the significance. These two values can be expressed from  
\begin{equation}
    S/B=\frac{\sigma_{S}}{\sigma_{B}},~~ S/\sqrt{B}=\frac{\sigma_{S}\times I}{\sqrt{\sigma_{B}\times I}},\label{equSignifice}
\end{equation}
where $\sigma_{S}$ represents the cross section of the signal and $\sigma_{B}$ represents the cross section of the backgrounds. $I$ represents the integrated luminosity of the collider, which is set as $1ab^{-1}$ in this work.

To distinguish the signal from the backgrounds, we firstly give some kinematic distributions in figure~\ref{B_cut}. The collision energy is 10 TeV and $m_{W^\prime}=2$ TeV. Here, $P_T$ represents the transverse momentum of the final state electrons in figure~\ref{B_cut} (a). $\eta$ is the pseudo-rapidity of electrons in figure~\ref{B_cut} (b). Another important parameter is $\alpha$ in figure~\ref{B_cut} (c), which is the azimuthal angle of the final state dilepton calculated with the angle between the two charged leptons in the transverse plane. And the last one in figure~\ref{B_cut} (d) is $\slashed{E}_T$, which represent the transverse energy carried away by the neutrinos. From these distributions we can find the discrepancy of the signal and background processes.

We attempted to the search of $W^\prime$ pair production with different masses of  2 TeV, 4.5 TeV and 4.8 TeV at a collision energy of 10 TeV. Different cuts are applied to enhance the significance. The results are listed in Table~\ref{tb4} in details. 
The basic cut is defined from the detector trigger constraints. We select the events with the number of final-state electrons to be two, and both of them being charge conjugates with the transverse momentum $P_T>10$ GeV.
When the cuts on the final-state electrons are set as $P_T>400$ GeV and $0.5< \alpha<3.1$, the signal-to-noise ratio can reach 17 and the significance can reach 32 for 2 TeV $W^\prime$. 
When the $W^\prime$ mass is up to 4.5 TeV, the cuts are changed to $600<P_T<3700$ GeV and $0.5< \alpha<3$.  The signal-to-noise ratio is 9.47 and the significance value is 9.14. 
The results of $m_{W^\prime} =4.8$ TeV is given with the cuts of $600<P_T<3500$ GeV and $0.5< \alpha<3$. The signal-to-noise ratio is 5.42 and the significance value is 5.17. With the mass of $W^\prime$ close to 5 TeV as the threshold of pair production for collision energy of 10 TeV, the cross section will decrease sharply with increasing mass, thus the significance will decrease obviously.


\begin{table}[H]
\begin{center}
\setlength{\abovecaptionskip}{6pt}
\setlength{\belowcaptionskip}{0pt}
\begin{tabular}{| c |c |c|c|}
\hline
\hline
 $m_{W^\prime}$ &cut range & $S/B$&$S/\sqrt{B}$ \\
\hline
\hline 
\multirow{3}{*}{2TeV}&Basic Cut & $0.0015$ &$0.31$\\
\cline{2-4}
\multirow{3}{*}{} &$P_T>400$GeV & $0.16$& $3.2$ \\
\cline{2-4}
\multirow{3}{*}{} &$0.5< \alpha<3.1$ & $17$ & $32$\\
\cline{1-4}
\multirow{3}{*}{4.5TeV}&Basic Cut & $2.1\times 10^{-4}$ &$0.047$\\
\cline{2-4}
\multirow{3}{*}{} &$600<P_T<3700$GeV & $0.066$& $0.78$ \\
\cline{2-4}
\multirow{3}{*}{} &$0.5< \alpha<3$ & $9.47$ & $9.14$\\
\cline{1-4}
\multirow{3}{*}{4.8TeV}&Basic Cut & $1.2\times 10^{-4}$ &$0.025$\\
\cline{2-4}
\multirow{3}{*}{} &$600<P_T<3500$GeV & $0.038$& $0.44$ \\
\cline{2-4}
\multirow{3}{*}{} &$0.5< \alpha<3$ & $5.42$ & $5.17$\\
\cline{1-4}
\hline
\end{tabular}
\caption{The signal-to-noise ratio and significance after cuts with  the mass of $W^\prime$ is 2, 4.5 and 4.8 TeV and the collision energy is 10 TeV.}
\label{tb4}
\end{center}
\end{table}

\section{Summary}

Extra charged gauge bosons beyond the Standard Model have always been a topic of interest. Future muon colliders, due to their high center-of-mass energies, have a significant advantage in discovering heavy particles. In this paper, we choose the ALRM model to study the $\mu^+  \mu^- \to W^{\prime +} W^{\prime -} \to e^+ e^- n_e \bar{n}_e$ process and explore the properties of $W^\prime$. In this process, we examine the cross section distribution, angular distribution, and forward-backward asymmetry for polarized and unpolarized scenarios of $W^\prime$ particles at different masses.

We have conducted an in-depth study of the cross section distribution of the process $\mu^+  \mu^- \to W^{\prime +} W^{\prime -} \to e^+ e^- n_e \bar{n}_e$ at different collision energies. 
The cross section can reach 0.2 $fb$ in a 10 TeV collider, which is
correlated to the combined effects of $g_R$ and the $W^\prime$ mass.
We also investigate the angle distributions of the initial-state muon and final-state electron for different $W^\prime$ masses. With a fixed $W^\prime$ mass, the asymmetry of the angular distribution gradually strengthens as the $g_R$ increasing.
With the same coupling strength, the changes in $W^\prime$ mass also greatly affect the angular distribution, but the overall trend remains consistent. We also investigate the cross section distribution and angular distribution for the polarized initial state conditions. When the $\mu^-$ polarization is different, the  distributions of cross section also show significant changes with the $W^\prime$ mass lower than half of the collision energy. But the polarization has slight effects on the angular distributions when the $W^\prime$ mass close to the threshold of pair production except for the case of $P_{\mu^+}$ extremely close to 100.
The forward-backward asymmetry defined from the angular distribution is an effective observable to study the interaction of $W^\prime$ coupling to leptons.
We also give the search of $W^\prime$ pair production at the muon collider with the mainly backgrounds of $ZZ$, $WW$ and VBF processes. After applying the suitable kinematic cuts, the significance can reach $5 \sigma$ for $W^\prime$ boson with mass of 4.8 TeV at the 10 TeV muon colliders.
The study of extra gauge particles is an important part of the new physics beyond standard model. Searches on the pair production of $W^\prime$ at the future muon collider provide new perspective for the new physics models. 

\section*{Ackonwledgement}
This work was supported by the Natural Science Foundation of Shandong Province under grant Nos.~ZR2022MA065 and ZR2024QA138.

\bibliographystyle{apsrev4-2}
\bibliography{Reference}
\end{document}